\documentclass[11pt]{revtex4}
\usepackage{graphicx}
\usepackage{psfig}
\newcommand{\be}{\begin{equation}}
\newcommand{\bea}{\begin{eqnarray}}
\newcommand{\ee}{\end{equation}}
\newcommand{\eea}{\end{eqnarray}}

\begin{document}

\title{Semiclassical Spectrum of Small Bose-Hubbard Chains: A Normal Form Approach}
\author{A.P. Itin$^{1,2}$ and  P. Schmelcher$^1$}
\affiliation{$^{1}$ Zentrum f\"{u}r optische Quantentechnologien,
Universit\"{a}t Hamburg, Luruper Chaussee 149, 22761 Hamburg, Germany\\
$^2$ Space Research Institute (IKI), Russian Academy of Sciences,
117997, 84/32 Profsoyuznaya Str, Moscow, Russia}

\begin{abstract}

We analyze the spectrum  of the 3-site Bose-Hubbard model with
periodic boundary conditions using a semiclassical method. The
Bohr-Sommerfeld quantization is applied to an effective classical
Hamiltonian which we derive using resonance normal form theory.
The derivation takes into account the 1:1 resonance between
frequencies of a linearized classical system, and brings nonlinear
terms into a corresponding normal form. The obtained expressions
reproduce the exact low-energy spectrum of the system remarkably
well even for a small number of particles $N$ corresponding to
fillings of just two particles per site. Such small fillings are
often used in current experiments, and it is inspiring to get
insight into this quantum regime using essentially classical
calculations.

\end{abstract}

\maketitle

\section{Introduction}

Recent experimental and theoretical progress in the field of
ultracold quantum gases has stimulated many studies at the
interface of traditionally different disciplines such as atomic
physics, quantum optics and condensed matter physics providing an
intriguing link to fundamental many-body problems
\cite{exp1,Bloch}. A specific, but yet particularly interesting
topic is the quantum-to-classical correspondence in degenerate
quantum gases such as the adiabaticity versus non-adiabaticity of
the quantum dynamics of finite matter-wave systems
\cite{RMP,Dziarmaga,Altland,Gurarie,Mathey,Grandi, Abdullaev,
Kamchatnov1, Kamchatnov2, Sacha, Ahufinger, ITorma}.

The Bose-Hubbard model (BHM) which we study here is a hallmark of
condensed matter theory, and was realized experimentally using
ultracold quantum gases in optical lattices \cite{Bloch} following
an ingenious theoretical suggestion \cite{Zoller}. At the same
time, in the classical limit it is described by the celebrated
Discrete Nonlinear Schrodinger Equation (DNLSE), which possesses
both a rich statics and  dynamics \cite{Eilbeck}. Properties of
the BHM at high fillings (many particles per site) are known to be
well-reproduced by several semiclassical methods, such as the
mean-field approximation \cite{Kevrekidis} or the truncated Wigner
approximation \cite{TWAZoller,TWAP}. At low fillings, one would
generally expect semiclassical methods to be inapplicable.
Interesting enough, in a two-site BHM it was shown recently
\cite{Korsch, Keeling} that semiclassical quantization reproduces
the quantum spectrum remarkably well. That is, when considering
the classical limit of the BHM and applying e.g. the
Bohr-Sommerfeld quantization to the classical action, one obtains
(semi)classical energies which show a good quantitative agreement
with the exact (quantum) energies. A similar approach has been
taken for the case of the three- and five-site BHM
\cite{Kolovsky}, with significant insights into the qualitative
properties of the quantum spectrum, but no explicit expressions
for the semiclassical spectrum having been obtained, and a
corresponding comparison of the quantum and semiclassical
predictions is still missing.

As we show here, the multi-site BHM exhibits an intricate
classical dynamics which renders the construction of an effective
Hamiltonian, necessary for a corresponding semiclassical
quantization, a nontrivial mathematical problem. Our approach for
the derivation of an effective classical Hamiltonian relies on
resonance normal form theory \cite{AKN}. We note that the
classical system linearized around its equilibrium can be
represented as a collection of harmonic oscillators, the
frequencies of these oscillators being doubly degenerate. A pair
of oscillators with frequencies in resonance 1:1 to each other can
be analyzed using normalization techniques. To this end one needs
to apply a series of canonical transformations that bring the
quadratic and quartic terms of the Hamiltonian to a normal form,
thereby completely eliminating all cubic terms. The obtained
normalized Hamiltonian, written in terms of action-angle
variables, depends on a pair of classical actions, which can then
be straightforwardly quantized.

We note that our use of the normalization technique is very
similar to that exploited recently in the studies of the
mean-field dynamics of a nonlinear Stimulated Raman Adiabatic
Passage (STIRAP) process \cite{PRL_STIRAP,PRA_STIRAP}. In its
simplest version, nonlinear STIRAP considers three uniform
condensates: an atomic Bose-Einstein condensate (BEC) in its
ground state and a molecular BEC in its ground and an excited
state coupled by a laser fields (a theory for the non-uniform case
has also been developed \cite{K_STIRAP}). Using a certain
time-dependent sequence of laser pulses, it is possible to convert
the atomic BEC into a ground-state molecular BEC without
populating the molecular BEC in the excited state. From a
mathematical point of view, the system is a Hamiltonian dynamical
system with two degrees of freedom possessing a rich dynamics,
including e.g. nonlinear instabilities due to 1:1 resonances. It
is an interesting fact that exactly the same degeneracy influences
the dynamics of small Bose-Hubbard chains, as we will show below.

In detail we proceed as follows. Section \ref{caqm} describes our
classical and quantum model. It contains a derivation of the
effective classical Hamiltonian, its semiclassical quantization
and a comparison with the exact numerical results for the spectra
for both the weak and strong interaction regime. Section \ref{cr}
provides our conclusions. The Appendix contains a discussion of
the two-site Bose-Hubbard chain, where the difference between weak
and strong interaction regimes becomes transparent.

\section{Classical and quantum model} \label{caqm}

Let us consider the 3-site Bose-Hubbard model with periodic
boundary conditions i.e. a ring geometry

\be H = -J \sum \limits_{\langle i,k \rangle} ( \hat
a_{i}^{\dagger} \hat a_k + h.c. ) + \frac{U}{2} \sum
\limits_{l=1}^3 \hat a_l^{\dagger} \hat a_l^{\dagger} \hat a_l
\hat a_l, \ee

where the first sum runs over all the nearest neighbours. As
discussed e.g. in Ref. \cite{Kolovsky}, one may try to understand
its spectrum using the quantization of a corresponding effective
classical Hamiltonian which will be derived in the following. To
apply a semiclassical quantization, we firstly  slightly modify
the BHM Hamiltonian, thereby writing it in a symmetrized form:

\be H = -J \sum \limits_{\langle i,k \rangle} ( \hat
a_{i}^{\dagger} \hat a_k + h.c. ) + \frac{U}{2} \sum
\limits_{l=1}^3 (\hat n_{l} + \frac{1}{2})^2,   \ee

where $\hat n_l = \hat a_l^{\dagger} \hat a_l$ is the particle
number operator for the $l-th$ site. Since the total number of
particles is constant, this modification introduces only a
(uniform) shift of all energy levels. The classical limit is then
obtained by introducing $c-numbers$ for the operators $\hat n_l$
in analogy to a classical coherent state formalism $\hat n_l
+\frac{1}{2} \to |\psi_l|^2$. This leads to the Discrete Nonlinear
Schr\"odinger (DNLS) equation

\be H = -J \sum \limits_{\langle i,k \rangle} (\psi_{i}^* \psi_k +
c.c.) + \frac{U}{2} \sum \limits_{l=1}^{3}  |\psi_l|^4 ,  \ee \bea
i \frac{d \psi_j}{dt} = \frac{\partial H}{\partial \psi_j^*} = -J(
\psi_{j+1} + \psi_{j-1} )  + U \psi_j |\psi_j|^2,
 \eea

It is important to note that the normalization of the classical
amplitudes is $ \sum_i |\psi_i|^2 = N_s = N + \frac{3}{2}$, where
$N$ is the total number of atoms in the quantum model
\cite{Korsch}. Introducing real pairs $(x_n,y_n)$ with $\psi_n=
\frac{x_n+i y_n}{\sqrt{2}}$, the DNLSE is equivalent to the
Hamiltonian equations of motion of the Hamiltonian \be H = U
\sum_{l=1}^3 \frac{(x_l^2 + y_l^2)^2}{8} - J \sum_{\langle i,k
\rangle} \left( x_i x_k + y_i y_k \right), \ee

Switching to polar coordinates $x_i = \sqrt{2n_i} \sin \phi_i,
\quad y_i = \sqrt{2 n_i} \cos \phi_i$, we arrive at the
Hamiltonian \be H = \frac{U}{2} \sum_{l=1}^3 n_l^2  - 2J
\sum_{\langle i,k \rangle} \left(\sqrt{n_i n_k} \cos(\phi_i -
\phi_k) \right), \ee

with $\sum_i n_i =\mbox{const}= N_s = N+\frac{3}{2}$. We now
perform a rescaling $n_i = I_i N_s$, $H = h N_s$. With $g=N_s U$
this leads to the Hamiltonian

\be h = \frac{g}{2} \sum_{l=1}^3 I_l^2 - 2J \sum_{\langle i,k
\rangle} \sqrt{I_i I_k} \cos (\phi_i - \phi_k), \ee

with $\sum_i I_i =\mbox{const}= 1.  \label{IphiHam}$

It is possible to introduce an effective classical Hamiltonian for
the low-energy and high-energy dynamics of the system. Here we
restrict ourselves to the low-energy part.

For $g>g_{cr}= -9/2$ the ground state is uniform in density and
phase $I_i=1/3 \hspace*{0.2cm} \forall i$ and $\phi_i=\phi_j
\hspace*{0.2cm} \forall i,j$. For strongly attractive interaction
$g<g_{cr}= -9/2$ the ground state is essentially different
\cite{Eilbeck}. Here we only consider values of interactions far
from this bifurcation, i.e. we have $ g > g_{cr} $.

\subsection{Low-energy spectrum for small and moderate values of interaction}

It is possible to eliminate one degree of freedom by applying a
transformation with the generating function $W = p_1(\phi_1 -
\phi_2) + p_2(\phi_1 +\phi_2-2\phi_3)/\sqrt{3} + p_3(\phi_1
+\phi_2+ \phi_3)/3$.

This generating function is chosen such that the expressions
occurring in the below-given Taylor expansion will have a simple
appearance.

The transformed Hamiltonian depends only on two new phases
$\theta_1, \theta_2$ and their corresponding momenta $p_1,p_2$,
while the third momentum is an integral of motion $p_3 =
\sum_{i=1}^3 I_i=1$. At values of $g$ larger than the critical
value $g_{cr}$, the stable equilibrium is at the origin
$p_{1,2}=0$, $\theta_{1,2}=0$. We expand our Hamiltonian around
the origin up to terms of fourth-order power in coordinates and
momenta. Without loss of generality we put $J=1$ in the following.

The resulting Hamiltonian is \be h = h_0 + \frac{\theta_1^2 +
\theta_2^2}{2} + \frac{9+2g}{2}(p_1^2 +p_2^2) + H_3 + H_4,  \ee

where $H_3$ and $H_4$ contain cubic and quartic terms,
respectively, and $h_0 = -2 +\frac{g}{6}$ is a constant. Note that
the quadratic part of the Hamiltonian describes two harmonic
oscillators whose frequencies are in resonance 1:1. It is exactly
this type of degeneracy which was recently considered in Ref.
\cite{PRL_STIRAP,PRA_STIRAP}.  The quadratic part of the
Hamiltonian implies a bifurcation at $g_{cr}=-9/2$, i.e. at
sufficiently strong attractive interaction. As already mentioned
above, we focus on the low-energy dynamics off this bifurcation,
i.e. for weakly attractive or repulsive interactions.

The cubic terms are given by

\be H_3= \frac{\sqrt{3}}{4}( 2 \theta_1 \theta_2 p_1 +
(\theta_1^2- \theta_2^2-27 p_1^2)p_2 + 9 p_2^3)
 \ee

To bring the Hamiltonian to its normal form, we need to get rid of
the cubic terms. This can be done by a nonlinear near-to-identity
canonical transformation $p_{1,2},\theta_{1,2} \to
P_{1,2},X_{1,2}$ determined by the generating function

\be W = \theta_1 (P_1 + \beta P_1 P_2) + \theta_2 (P_2 + \alpha
P_2^2 + \gamma P_1 ^2 + c \theta_1^2 + d \theta_2^2), \ee

with the coefficients \bea \alpha = -\frac{9 \sqrt{3}}{4A}, \quad
\beta= - 2 \alpha, \quad \gamma = -\alpha,
\\ c = \sqrt{3}\frac{18-A}{4A^2}, \quad d = -c/3, \quad   A= 9+2g \eea

This transformation does not change the quadratic terms, removes
the cubic terms, and modifies the quartic ones. We subsequently
change to polar coordinates $J_{1,2},\Phi_{1,2}$, with
$X_{1,2}=\sqrt{2 J_{1,2}} A^{1/4} \sin{\Phi_{1,2}},
P_{1,2}=\sqrt{2 J_{1,2}} A^{-1/4} \cos{\Phi_{1,2}}$

In the resulting Hamiltonian,  we keep only slowly varying terms
which depend on $\Phi_1-\Phi_2$, and average out (i.e. omit) other
('fast') trigonometric terms, e.g. $\cos 2 \Phi_1, \cos 2\Phi_2,
\cos(2(\Phi_1+\Phi_2))$, etc. We thus arrive to the normal form

\be H =  \sqrt{A} (J_1 +J_2) + B(J_1^2 + J_2^2) + C J_1 J_2 + D
J_1 J_2 \cos [2(\Phi_1-\Phi_2)], \ee

where

\bea B &=& -\frac{3g}{8A^2}(54+24g+g^2), \quad C = -
\frac{g}{2A^2}(6+g)(18+g),
\\ D &=& -\frac{g}{4A^2}(-54+24g+g^2),  \eea

One can easily see that at $g=0$ all coefficients $B,C,D$ are
equal to zero, implying the degeneracy of the Hamiltonian with
respect to $J_1-J_2$ at this point. Introducing
$J=\frac{J_1+J_2}{2}, P=\frac{J_1-J_2}{2}$, we get the Hamiltonian

\be h = 2 \sqrt{A} J +2B (J^2 + P^2) + C(J^2-P^2) + D
(J^2-P^2)\cos 2\Phi \ee

where $P$ and $\Phi$ are canonically conjugate. This Hamiltonian
is integrable, since $J=$const. $J$ is the first action variable
of our effective Hamiltonian, the second one we find by
calculating  $K= \frac{1}{2 \pi} \int P d\Phi$.  This integral can
be calculated analytically, which provides us with an expression
for $K(h)$. Inverting it, we finally find that

\be h =  2 \Omega J + c J^2 - e [K^2 - 2KJ], \label{h3classical}
\ee
 where
\be c = 2C= -\frac{g}{A^2}(6+g)(18+g), \quad e = D =
-\frac{g}{4A^2} (-54+24g+g^2), \quad \Omega=\sqrt{A},
\label{constceomega} \ee

The above expression (\ref{h3classical}) together with the
constants (\ref{constceomega}) constitute the main result of this
paper. The quantization of the actions $J,K$ should reproduce the
low-lying energy levels of the system. We use the following
quantization of the actions
\bea 2J &=& \frac{n+1}{N_s}, \quad n = 0,1,.. N,  \\
    2K &=& \frac{m+1/2}{N_s},  \quad m =0,..n,
\eea

\begin{figure}
\includegraphics[width=90mm]{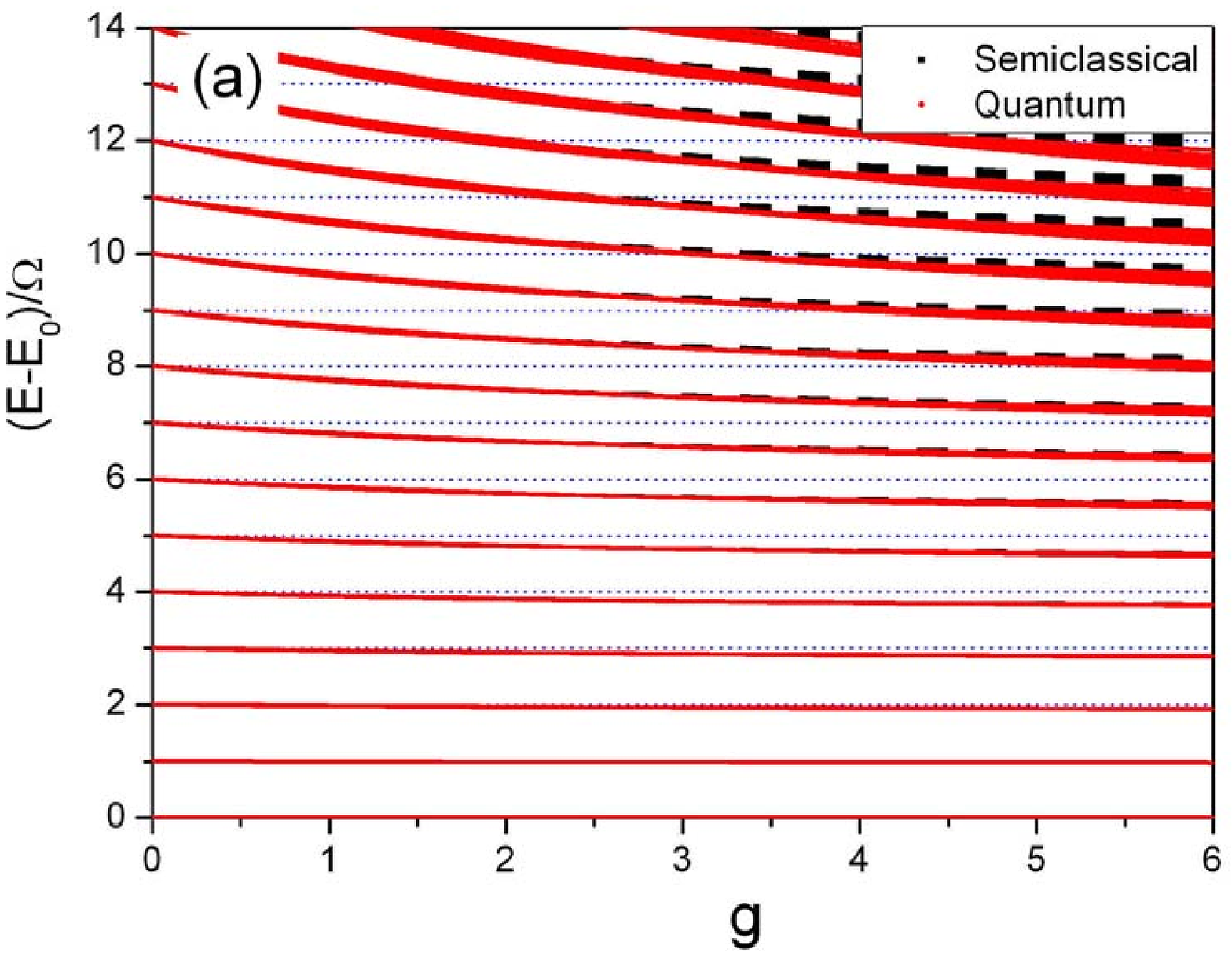}
\includegraphics[width=90mm]{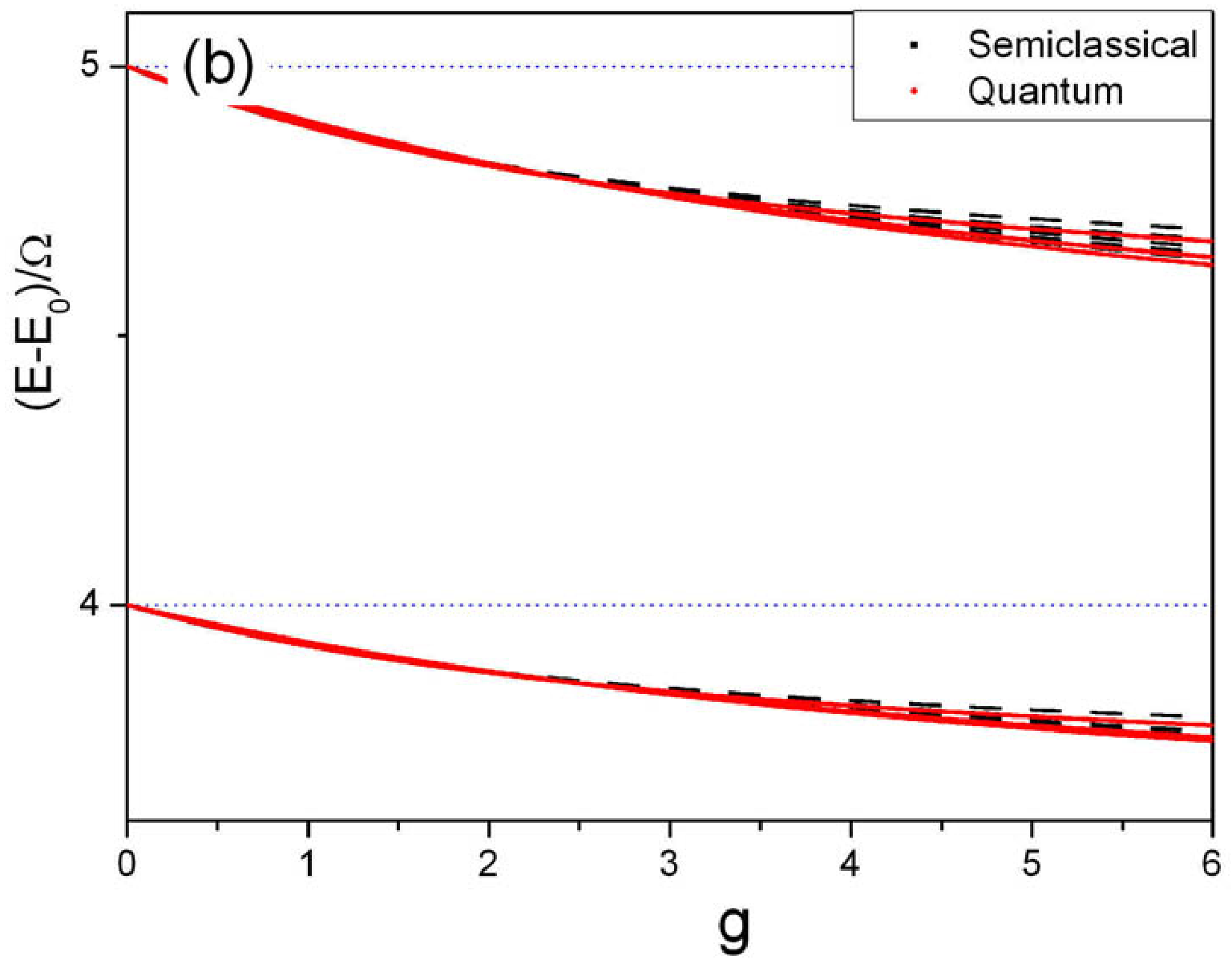}
\caption{Energy levels of the three-site BHM (red solid lines) and
semiclassical levels given by Eq. (\ref{h3classical}) (black
dashed lines) as a function of the interaction strength $g$ for
$N=24$ atoms. The energy levels are rescaled to the Bogoliubov
frequencies, i.e. we show $(E-E_0)/\Omega$, where $E_0$ is the
energy of the ground state, and $\Omega=\sqrt{A}=\sqrt{9+2g}$. The
Bogoliubov levels are denoted by blue dotted lines.
 } \label{quantumclassical3}
\end{figure}

\begin{figure}
\includegraphics[width=90mm]{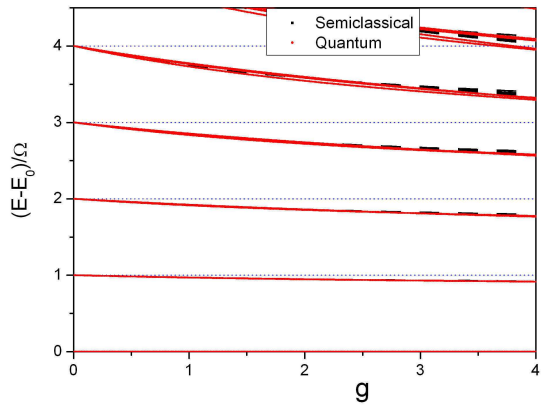}
\caption{Energy levels of the three-site BHM (red solid lines) and
semiclassical levels given by Eq. (\ref{h3classical}) (black
dashed line) as a function of the interaction strength $g$ for
$N=6$. The energy levels are rescaled to the Bogoliubov
frequencies. The corresponding Bogoliubov levels are denoted by
blue dotted lines.
 } \label{N6quantumclassical}
\end{figure}

The resulting semiclassical energy-levels $E_{nm}$ are shown in
Figs. \ref{quantumclassical3}, \ref{N6quantumclassical},  in
comparison with the results from exact numerical diagonalization
for N=24 and N=6. It is observed that for moderate values of the
interaction strength $|g|$ the exact spectrum is reproduced very
well. To be more specific, there are two phenomena occuring for
the spectrum with increasing $|g|$. At zero interaction the exact
spectrum coincides with the Bogoliubov one, and the energy levels
are organized in (degenerate) Bogoliubov bands. As $g$ is
increased, these energy bands decrease and their degeneracy is
lifted.

The first phenomenon is reproduced by the semiclassical approach
remarkably well. That is, we see from
Figs.\ref{quantumclassical3}, \ref{N6quantumclassical} that the
energetical distance between the exact solutions and the
semiclassical prediction is much less than the distance between
the exact and the linearized solutions (i.e., the deviation of the
exact spectrum from the Bogoliubov frequencies is much larger than
its deviation from the semiclassical prediction). At the same
time, the spreading of the energy levels within the Bogoliubov
band is reproduced not very well. For large $g$-values the
deviation is considerable.  As $g$ is increased, fluctuations with
respect to the phase grow and an expansion around the classical
ground state becomes inapplicable: the phase is not confined to
the vicinity of 0 anymore. The case of large $g$ is analyzed in
the next section.

\subsection{Low-energy spectrum for strong interactions}

At strong interactions, even though the {\em classical} ground
state remains unchanged (uniform in density and phase), the {\em
semiclassical} ground state becomes qualitatively different. The
semiclassical ground state includes zero-point oscillations, i.e.
the corresponding classical trajectories possess certain non-zero
classical actions. With increasing interaction strength, the area
of the phase space filled with trajectories oscillating around the
classical ground state shrinks. At a certain value of $g$ the
oscillatory area cannot accommodate trajectories with these
minimal actions. As a result, the nature of the classical motion
changes from oscillatory to rotational.  In other words, the
zero-point oscillations destroy the phase coherence. This is most
easily illustrated for the double-well case (see supplementary
information). One can consider it as a (semi)classical counterpart
of the Mott-insulator transition. To be more precise, in finite
quantum chains the Mott-insulator transition becomes a smooth
crossover, and the change of the classical motion described above
is the classical counterpart of this {\em crossover}.

From a technical point of view, this qualitative change in
classical trajectories corresponding to the ground and low-lying
states leads to the inapplicability of the fourth-order expansion
of the classical Hamiltonian around the origin: phases now cover
the interval $[-\pi,\pi)$ and are not restricted to the vicinity
of zero. An adequate effective Hamiltonian will consequently
change severely. To derive it, one can use classical perturbation
theory such as Linstedt's method \cite{AKN}.

Similar to the case of weak interactions, we introduce \be P_1 =
\frac{I_1-I_2}{2}, \quad P_2 =\frac{1}{6}(I_1+I_2-2I_3), \quad P_3
= \sum \limits_{i=1}^{3} I_i =1, \ee

Expanding the Hamiltonian in powers of $P_1,P_2$ and $1/g$, one
obtains \be H= g[ \frac{1}{6} + P_1^2 + 3P_2^2 - \frac{2}{3g}(\cos
\Phi_1 +2\cos \frac{\Phi_1}{2} \cos \frac{\Phi_2}{2} ) + O(P_i/g)
]  \ee

We assume that the interaction strength $g$ is large enough such
that $P_i>\sqrt{1/g}$ holds. This holds if $1/N \gg \sqrt{1/g}$.
Then, the classical dynamics takes place outside the resonance
zones i.e. it becomes a rotational motion. Then, in the first
approximation of Linstedt's method, one introduces new actions
$P_1 \to J_1, P_2 \to J_2$, and the averaged Hamiltonian in the
new actions $J_1,J_2$ coincides with the unperturbed initial
Hamiltonian (this result can be obtained by other methods as well
\cite{Benetin}), \be H= g[ \frac{1}{6} + J_1^2 + 3J_2^2  ]
 \ee

The semiclassical energy levels are obtained by  quantizing
$J_1,J_2$: \be E_{n_1,n_2} = gN_s (\frac{1}{6} + j_1^2 + j_2^2 ),
j_1= \frac{n_1-n_2}{2N_s}, \quad
j_2=\frac{n_1+n_2}{2N_s}-\frac{1}{3}, \quad n_1=0,..,N, \quad n_2=
0,..,N-n_1, \label{LGclassical} \ee

\begin{figure}
\includegraphics[width=100mm]{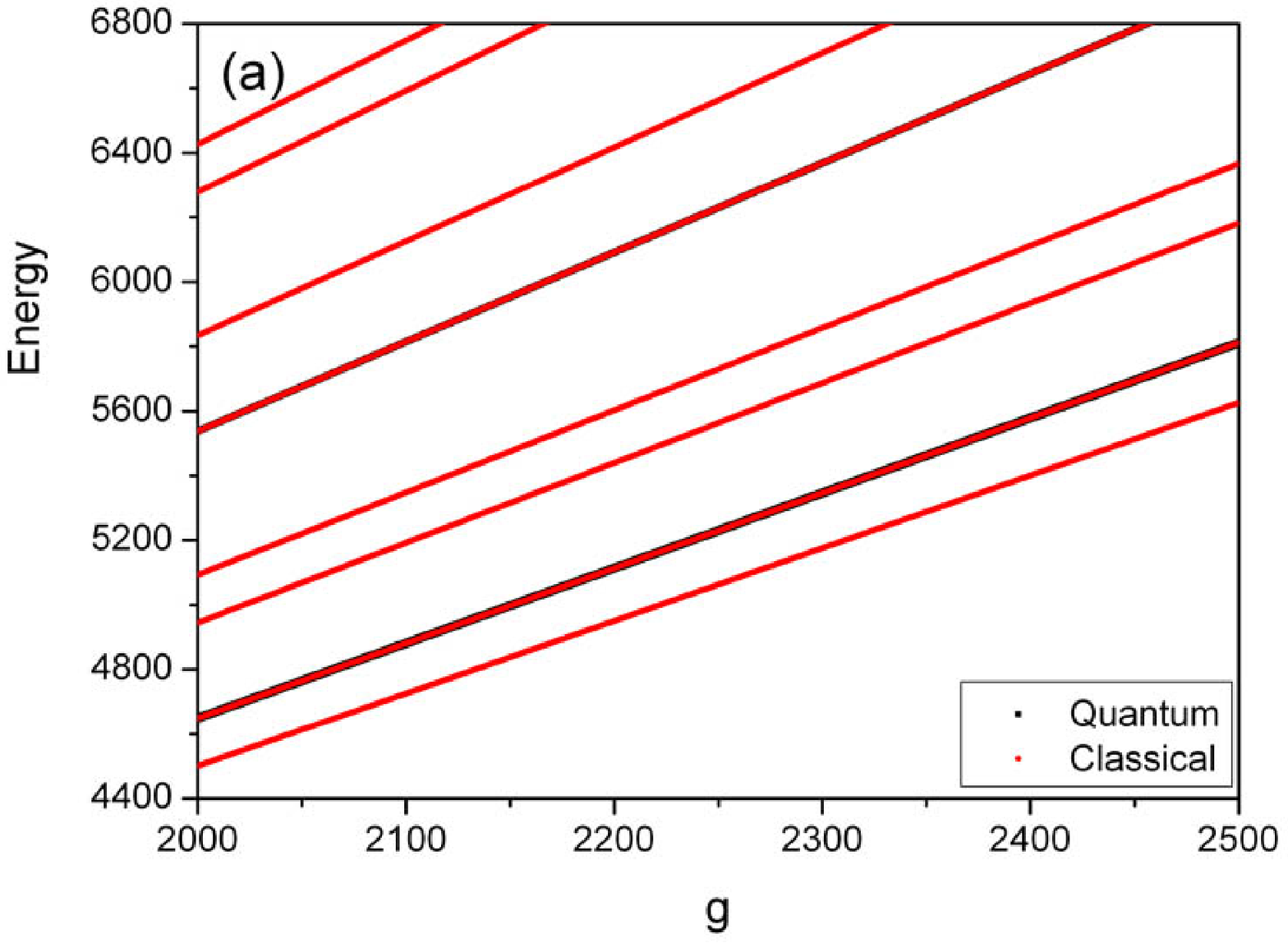}
\includegraphics[width=100mm]{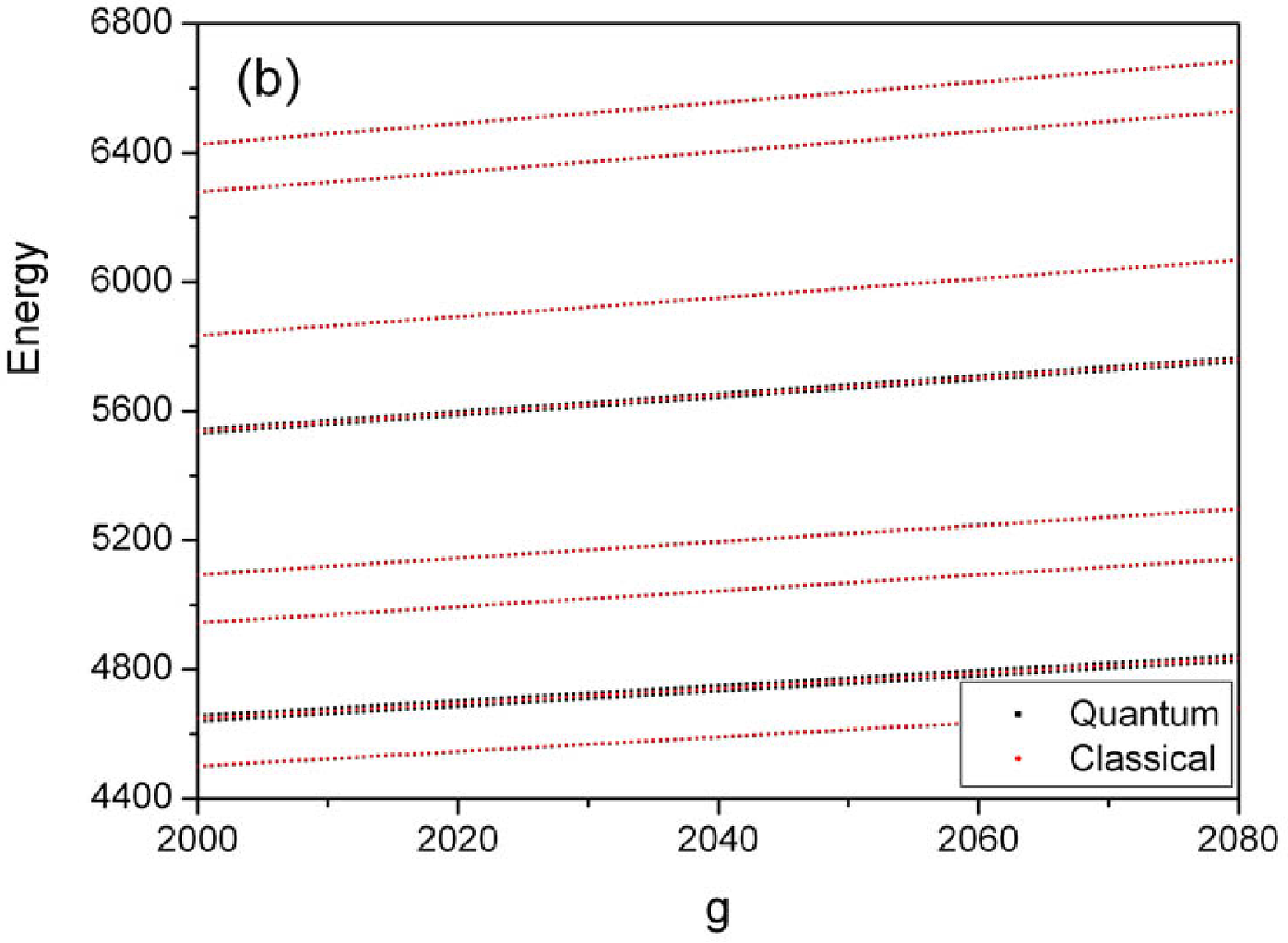}
\caption{Energy levels of the three-site BHM (black squares) and
semiclassical energy levels given by Eq. (\ref{LGclassical}) (red
circles) as a function of the interaction strength $g$. The levels
are almost indistinguishable on the scale of (a). (b) represents
an enlarged view of a part of (a).
 } \label{N7quantumclassicalLG}

\end{figure}

Fig.\ref{N7quantumclassicalLG} shows a comparison of the
semiclassical and exact results. In the second approximation of
Lindstedt's method, the dependence of the new Hamiltonian on the
actions is slightly modified and the corresponding expressions are
more cumbersome and therefore not provided here.

\section{Concluding remarks} \label{cr}

We have shown that by performing a proper analysis of the
classical counterpart of a few-site Bose-Hubbard model one can
gain valuable insights into the dynamics of the system.

The classical system possesses two degrees of freedom and its
small-amplitude oscillations around the ground (equilibrium) state
can be brought into the form of two decoupled linear oscillators.
An important feature of the system is that the oscillators are in
resonance 1:1 to each other. As a subsequent approximation,
considering excitations on top of the ground state with larger
amplitude, one should therefore apply resonance normal form
theory. This allows us to bring the system to the form of an
integrable nonlinear Hamiltonian depending on two classical
actions. The quantum spectrum can then be reproduced by quantizing
these actions. From the point of view of the physics of
Bose-Einstein condensates, this procedure can be seen as an
extension of the Bogoliubov transformations to the realm of
nonlinear oscillations.

For strong interactions, the properties of the system change
drastically. Here perturbation theory with respect to the inverse
interaction strength leads to a system of uncoupled rotators.

An interesting dynamics occurs if one dynamically sweeps the
interaction from strong to  weak values (or vice versa), passing
through the crossover region. Depending on the sweeping rate of
the interaction, the crossover region is passed adiabatically or
non-adiabatically and correspondingly different amounts of
excitations are produced at the end of the sweep. This problem for
the triple-well is left for future research. However, we solved it
for the double-well. This allows, e.g, to construct a mathematical
theory of slow decoupling of two superfluids, extending the
results of \cite{Mathey} on abrupt decoupling of superfluids to
the case of slow sweeps. This question is presented elsewhere
\cite{Schmelcher}.


The results can be generalized to the case of longer chains. While
it is understood that normal form theory is useful for weakly
interacting wave dynamics \cite{Lvov}, we are not aware about
corresponding explicit calculations in BHM.

\section*{Acknowledgments}

We thank A.I.Neishtadt, P.Kevrekidis and F.K.Diakonos for useful
discussions. A.P.I. acknowledges partial support from the RFBR
grant 09-01-00333.

\appendix

\section{Quantization of the two-site chain}

For comparison, we consider here the quantization in case of the
double-well problem. The two-site Bose-Hubbard Hamiltonian reads

\be \hat H= -J(\hat a_2^{\dagger} \hat  a_1 + \hat  a_1^{\dagger}
\hat  a_2  ) + \frac{U}{2} [\hat  n_1(\hat n_1-1) +\hat n_2(\hat
n_2-1) ] \ee

The symmetrized form of the interaction term  is \be \frac{U}{2}[
(\hat n_1+\frac{1}{2})^2 + (\hat n_2+\frac{1}{2})^2 -\frac{1}{2} -
2(\hat n_1 + \hat n_2) ], \ee

therefore we will consider below the slightly modified form of the
BHM, additionally asuming $J=1$:

\be \hat H= -(\hat a_2^{\dagger} \hat  a_1 + \hat  a_1^{\dagger}
\hat  a_2  ) + \frac{U}{2} [ (\hat n_1+ \frac{1}{2})^2 +\hat (\hat
n_2 + \frac{1}{2})^2 ] \ee The semiclassical limit is given by the
Hamiltonian \be H = -( \psi_2^* \psi_1 + \psi_1^* \psi_2) +
\frac{U}{2} [|\psi_1|^4 + |\psi_2|^4 ], \ee with the corresponding
equations of motion \be i \frac{d \psi_j}{dt}= \frac{\partial
H}{\partial \psi_j^*}, \quad  i \frac{d \psi_j^*}{dt}=
-\frac{\partial H}{\partial \psi_j}. \ee

The normalization of the semiclassical variables is \cite{Korsch}
\be |\psi_1|^2 + |\psi_2|^2 = N+1 = N_S \ee Introducing $\psi_i =
\frac{x_i + y_i}{\sqrt{2}}$, the equations of motion for $x_i,y_i$
are determined by the classical Hamiltonian \be H = \frac{U}{8} [
(x_1^2 + y_1^2)^2 + (x_2^2 + y_2^2)^2 ] -(x_1 x_2 + y_1 y_2),
\label{classical} \ee

which after transforming $x_i = \sqrt{2 I_i} \cos \phi_i, \quad
y_i = \sqrt{2 I_i} \sin \phi_i$ becomes \be H = \frac{U}{2}(I_1^2
+I_2^2) - 2 \sqrt{I_1 I_2} \cos{(\phi_1-\phi_2)}
\label{classical2} \ee

with the integral of motion $I_1+I_2 = N_s$. Employing the
generating function $W = \frac{P}{2}(\phi_1-\phi_2) +
\frac{N_s}{2}(\phi_1 + \phi_2)$ to canonically transform the above
Hamiltonian we arrive at \be H = \frac{U}{4} (P^2 + N_s^2) -
\sqrt{N_s^2-P^2} \cos 2 \phi \ee

Subsequently the transformation $P= N_s p$, $ H = N_s h$, and
gives us \be h =  \frac{g p^2}{4}  - \sqrt{1-p^2} \cos 2 \phi, \ee
where $ g = U N_s$. For weak interactions, a phase point
oscillates around the ground state.  For strong interactions,
separatrix $h=$ shrinks to the vicinity of $p=0$,  and when the
area within the separatrix divided by $2 \pi$ becomes less than
the minimal action $\frac{1}{2 N_s}$, nature of motion of the
phase point corresponding to the semiclassical ground state
changes: it is a rotation, with $\phi$ covering the full interval
$(0,2\pi)$.

While it is not difficult to calculate the classical action here,
we would like to follow a general scheme which could be
generalized to a $M-site$ Bose-Hubbard chain. Therefore for weak
interactions we proceed with an expansion of the Hamiltonian
around its equilibrium, keeping terms of up to the fourth order.
Expanding the Hamiltonian close to the origin, one gets \be h = -1
+ 2 \phi^2+ p^2 (1+g/2)/2 +\frac{p^4}{8} -p^2 \phi^2 -\frac{2
\phi^4}{3} \ee

The quadratic part can be transformed to action-angle variables
easily \be p=\sqrt{2I/\omega} \cos x, \quad \phi = \sqrt{2I
\omega} \sin x, \quad \omega = \frac{1}{2}\sqrt{1+g/2}
 \ee

Then we obtain \be h  =-1+ \Omega I  - \frac{g(8+g)}{4 \Omega^2}
I^2 + I^2 F(x), \ee where $\int \limits_{ x =0}^{2\pi} F(x) = 0$,
$\Omega = 4 \omega = \sqrt{2(2+g)}.$ Averaging over $x$, we get
the effective Hamiltonian \be h  =-1+ \Omega I  - \frac{g(8+g)}{4
\Omega^2} I^2 \ee

\begin{figure}
\includegraphics[width=90mm]{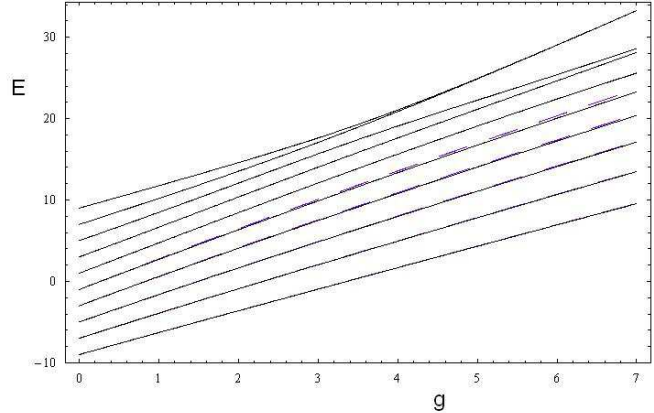}
\caption{Energy levels of the two-site BHM (black solid lines) and
the first 5 semiclassical levels (\ref{semiclassical}) (blue
dashed lines) for $N=9$ atoms. The first three levels are almost
indiscernible on this scale.} \label{quantumclassical}
\end{figure}

The quantization of the action leads to $I_n = \frac{n+1/2}{N_s},
\quad n=0,..N $ and finally gives us the semiclassical energy
levels \be E = N_s \left( \frac{g}{4} - 1 + \Omega I_n -
\frac{g(g+8) I_n^2}{4 \Omega^2} \right), \label{semiclassical} \ee

which reproduce the low-energy spectrum of the dimer BHM
remarkably well even for $N \sim 10$ (see Fig.
\ref{quantumclassical}).

\begin{thebibliography}{10}

\bibitem{exp1} I. Bloch, J. Dalibard, and W. Zwerger,
Rev. Mod. Phys. {\bf 80}, 885 (2008); S.\ Giorgini, L.P.\
Pitaevskii, and S.\ Stringari, {\it ibid} {\bf 80}, 1215 (2008).

\bibitem{Bloch} M. Greiner, O. Mandel, T. Esslinger, T.W. Hansch, and I.
Bloch, Nature {\bf 415}, 39 (2002).

\bibitem{Dziarmaga} Dziarmaga, Adv. Phys. {\bf 59}, 1063 (2010).

\bibitem{RMP} A. Polkovnikov, L.Sengupta, A.Silva, M.Vengalattore,
Rev. Mod. Phys. {\bf 83}, 863 (2011).


\bibitem{Altland} A. Altland {\it et al.},
Phys.Rev. A {\bf 79}, 042703 (2009).

\bibitem{Gurarie} V. Gurarie,  Phys. Rev. A {\bf  80}, 023626 (2009).

\bibitem{Mathey} L.Mathey, A.Polkovnikov,  Phys. Rev. A {\bf 81}, 033605 (2010).

\bibitem{Grandi}  C.De Grandi, A. Polkovnikov, in Lect. Notes Phys. {\bf 802},  Eds.
A. Chandra, B.K. Chakrabarti,  Springer, Heidelberg, 2010.


\bibitem{Abdullaev} F. Kh. Abdullaev et al, Phys. Rev. A {\bf  67},
013605 (2003).

\bibitem{Kamchatnov1} F.Kh. Abdullaev, A.M. Kamchatnov, V.V.
Konotop, and V.A. Brazhnyi, Phys. Rev. Lett {\bf 90}, 230402
(2003).

\bibitem{Kamchatnov2} T.L. Horng et al, Phys. Rev. A {\bf 79}, 053619
(2009).

\bibitem{Sacha} B.Damski, K.Sacha, J. Zakrzewski, J.Phys. B {\bf
35}, 4051 (2002).

\bibitem{Ahufinger}  C.Ottaviani, V.Ahufinger, R.Corbaan, J. Mompart, Phys. Rev A
{\bf 81}, 043621 (2010)

\bibitem{ITorma}
 A.P.\ Itin, P.T\"{o}rm\"{a}, arXiv::0901.4778 (2010);
 A.P.\ Itin, P. T\"{o}rm\"{a}, Phys. Rev. A {\bf  79}, 055602
(2009).

\bibitem{Zoller}
D. Jaksch, C. Bruder, J. I. Cirac, C.W. Gardiner, and P. Zoller,
Phys. Rev. Lett. {\bf  81}, 3108 (1998).

\bibitem{Eilbeck}
J.C. Eilbeck, P.S. Lomdahl and A.C. Scott, Physica D {\bf 16}, 318
(1985).

\bibitem{Kevrekidis} A. Smerzi, A. Trombettoni, P.G. Kevrekidis, A.R.
Bishop, Phys. Rev. Lett. {\bf 89}, 170402 (2002).

\bibitem{TWAZoller} C.W. Gardiner and P. Zoller, Quantum Noise (Springer-Verlag,
2004).

\bibitem{TWAP}  A.Polkovnikov, Phys. Rev. A {\bf  68}, 033609 (2003).

\bibitem{Korsch}  E.M. Graefe anf H.J. Korsch, Phys. Rev. A {\bf
76}, 032116 (2007).

\bibitem{Keeling} F. Nissen, J.Keeling,  Phys. Rev. A {\bf 81}, 063628
(2010).

\bibitem{Kolovsky} A.R. Kolovsky, Phys. Rev. Lett {\bf 99}, 020401
(2007); A.R. Kolovsky, Phys. Rev. E {\bf 76}, 026207 (2007).

\bibitem{AKN} V.I.\ Arnold, V.V.\ Kozlov, and A.I.\ Neishtadt,
 Mathematical aspects of classical and celestial mechanics
  (Third Edition, Springer, Berlin, 2006).

\bibitem{PRL_STIRAP} A.P.Itin, S.Watanabe, Phys. Rev. Lett. {\bf
99}, 223903 (2007).
\bibitem{PRA_STIRAP} A.P.Itin, S.Watanabe, and V.V. Konotop,
Phys.Rev. A {\bf 77}, 043610 (2008).

\bibitem{K_STIRAP} H.A. Cruz, V.V. Konotop, Phys. Rev. A {\bf 83}, 033603 (2011).
\bibitem{Benetin} G.Benettin, L. Galgani, A. Giorgilli, Nuovo
Cimento B {\bf 89}, 89 (1985).

\bibitem{Schmelcher} A.P.Itin et al, in preparation.

\bibitem{Lvov} V.S. L'vov,  Wave turbulence under parametric
excitation (Springer-Verlag, Berlin, 1994),  chapter 1.

\end{thebibliography}
\end{document}